\documentclass[sigconf]{acmart}
\usepackage{threeparttable}
\usepackage{multirow}
\usepackage{pgfplots}
\usepackage{pgfplotstable}
\usepackage[normalem]{ulem}
\usepackage{xcolor} 
\pgfplotsset{compat=1.18}

\usepackage{tabularx}
\usepackage{subcaption}
\usepackage{longtable} 
\usepackage{booktabs}
\usepackage{caption}      
\usepackage{adjustbox}
\usepackage{tikz}
\usepackage{booktabs,array}
\usepackage{amsmath}        
\usepackage{mathtools}      
\usepackage{algorithm}
\usepackage{algpseudocode}
\usepackage{graphicx}       
\usetikzlibrary{arrows.meta,positioning,fit,shapes.misc}

\makeatletter
\renewcommand{\sout}[1]{\@gobble{#1}}
\makeatother

\AtBeginDocument{%
  }

\setcopyright{acmlicensed}
\copyrightyear{2026}
\acmYear{2026}
\acmDOI{10.1145/3830398.3830708}

\acmConference[UIST '26]{The 39th Annual ACM Symposium on User Interface
  Software and Technology}{November 2--5, 2026}{Detroit, MI, USA}

\acmISBN{}

\begin{document}
\raggedbottom

\title{AdaptaFace: Adversarial Assistance for Inclusive Facial Gesture Interaction}

\author{Siyu Zhang}
\email{s.zhang@bristol.ac.uk}
\orcid{0009-0003-2754-5529}
\affiliation{%
  \institution{University of Bristol}
  \city{Bristol}
  \country{United Kingdom}
}

\author{Yelu Gu}
\email{yelu.gu@bristol.ac.uk}
\orcid{0009-0005-5317-9467}
\affiliation{%
  \institution{University of Bristol}
  \city{Bristol}
  \country{United Kingdom}
}

\author{Xiaoyun Xu}
\authornote{Corresponding author.}
\email{xuxiaoyun@bza.edu.cn}
\orcid{0000-0001-5726-163X}
\affiliation{%
  \institution{Zhongguancun Academy}
  \city{Beijing}
  \country{China}}

\author{Kirsten Cater}
\email{kirsten.cater@bristol.ac.uk}
\orcid{0000-0002-3435-2392}
\affiliation{%
  \institution{University of Bristol}
  \city{Bristol}
  \country{United Kingdom}}

\author{Oussama Metatla}
\email{o.metatla@bristol.ac.uk}
\orcid{0000-0002-3882-3243}
\affiliation{%
  \institution{University of Bristol}
  \city{Bristol}
  \country{United Kingdom}}

\renewcommand{\shortauthors}{Zhang et al.}

\begin{abstract}

Facial gestures can enable access for people with motor impairments (PwM), yet recognizers may not reliably interpret individual facial movement patterns. Existing approaches either constrain agency by requiring predefined templates or impose repeated calibration, data collection, and computation through user-specific retraining. We present AdaptaFace, an input-level compatibility layer using bounded, proxy-dependent adversarial perturbations to translate user-compatible gestures into inputs for an unchanged facial action unit (AU)-based recognizer. Designed around six personalization principles informed by interviews with 12 PwM, AdaptaFace supports both recognition enhancement and gesture vocabulary extension. In a within-subject study with 16 participants, AdaptaFace increased task completion from 59.7\% to 90.5\% and reduced workload from 5.61 to 4.21. Offline validation showed strong steering with aligned perturbation generation and evaluation but weaker cross-recognizer transfer, defining the approach’s boundary. These findings frame adversarial assistance as input-side accessibility mediation for personalized facial-gesture interaction.

\end{abstract}

\begin{teaserfigure}
\centering
  \includegraphics[width=0.95\textwidth]{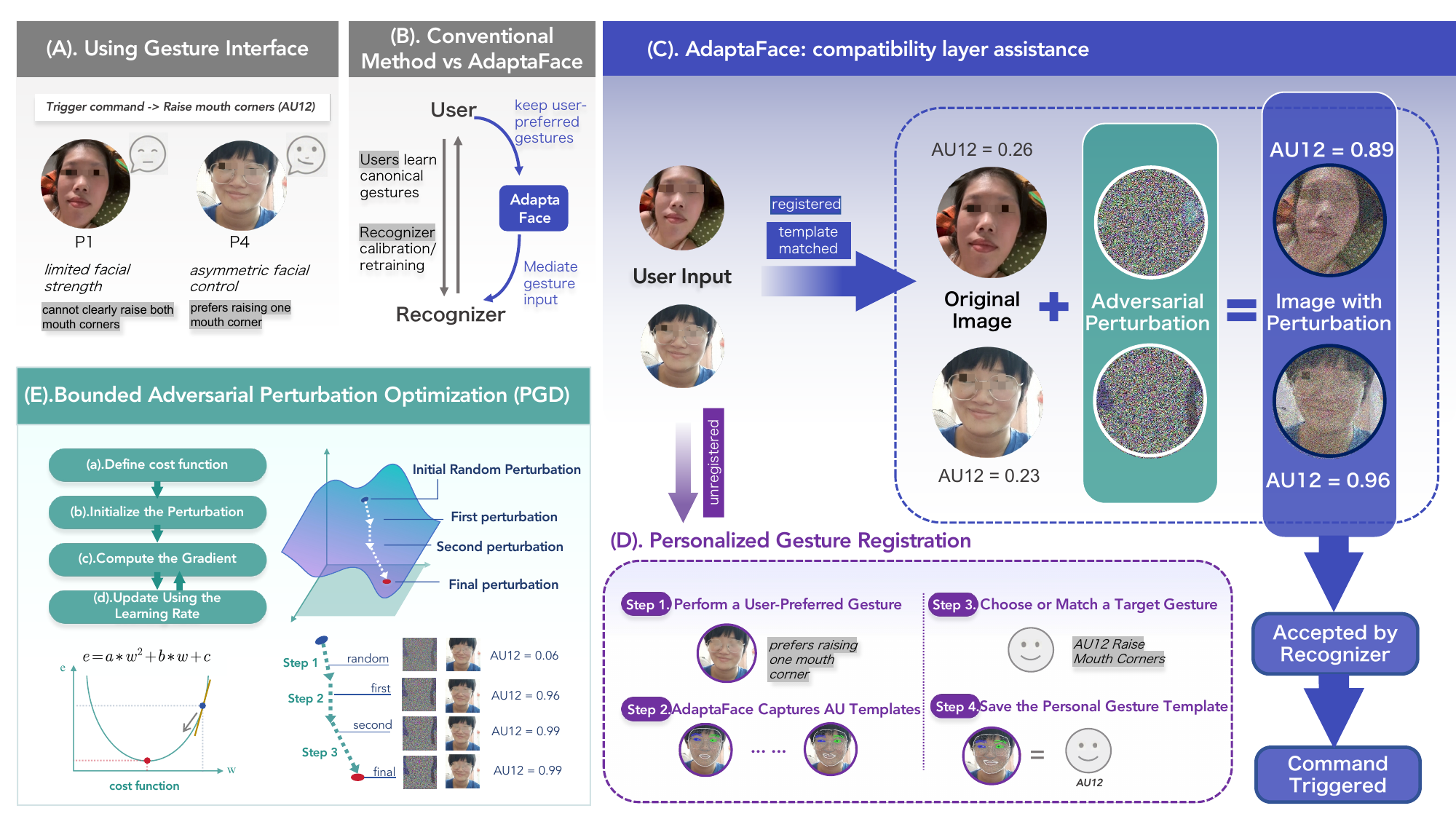}
  \caption{AdaptaFace overview: (A) user--recognizer mismatch, (B) input-side mediation, (C) runtime adversarial assistance, (D) personalized gesture registration, and (E) bounded perturbation optimization. AU12 denotes the lip corner puller.}
  \Description{AdaptaFace registers a user's feasible facial movement, associates it with an action-unit target, and applies a bounded perturbation to each recognized input frame before an unchanged recognizer processes it. The figure contrasts this input-side mediation with user adaptation and recognizer retraining and illustrates the iterative perturbation process.}
  \label{fig:core_mechanism}
\end{teaserfigure}




\maketitle

\section{Introduction}
\label{sec:introduction}

Facial gestures such as head movements, blinks, and mouth actions provide an important access pathway for people with motor impairments (PwM) when hand-based input is difficult or impossible~\cite{Betke2002CameraMouse,Wobbrock2011ABD,Zhao2022AboveNeck}. Because they can be captured using commodity cameras and require little or no upper-limb movement, they have long been seen as a promising modality for accessible interaction.
Yet, facial-gesture interfaces commonly expose a small set of canonical movements. Users whose facial movements are subtle, asymmetric, personalized, or variable may not align reliably with those predefined gestures~\cite{Zhao2022AboveNeck,Zhang2026BeyondAccuracy}. When recognition depends on reproducing canonical expressions, many PwM must exaggerate, repeat, or alter their movements to make themselves legible to the recognizer. This shifts the burden of adaptation onto users, undermining both reliability and agency in everyday interaction~\cite{Wobbrock2011ABD,CostanzaChock2020DesignJustice,Bates2003EyeMice,Namnakani2023CHI}. Interaction interfaces built on such recognizers can therefore produce allocative harms for PwM~\cite{Zhang2026BeyondAccuracy}.

Prior personalization strategies commonly modify interaction mappings or adapt recognizers using user-specific data. Mapping approaches are bounded by the configurations an implementation exposes, while model adaptation requires calibration samples or demonstrations~\cite{Gajos2010Supple,Wentzel2025MotionBlocks,He2019SAGE,Park2019FAZEICCV,Shahi2024SingleDemoCustomization}. This motivates a third path that improves compatibility without changing model parameters or requiring users to conform to canonical gestures.

We investigate this third path through a compatibility layer that mediates between user expression and recognizer expectations. We operationalize this logic through adversarial assistance, repositioning adversarial methods not as threats to recognition systems, but as a mediation mechanism for accessible interaction. AdaptaFace uses a differentiable proxy to compute subtle perturbations to live facial input so that an unchanged AU-based recognizer is more likely to interpret gestures that match a user's own motor repertoire. The recognizer is not retrained, but the perturbation remains dependent on the proxy and target output space. This reframes personalization from requiring users to conform to recognizer-defined templates or retraining recognizers for each user toward connecting the two through a mediating layer.

We developed and evaluated AdaptaFace in three stages. We interviewed 12 PwM to derive six design principles for low-burden personalization. We then built a browser-based system with gesture registration, assistance tuning, and four personalization modes: Fixed Enhancement, Matched Enhancement, Fixed Extension, and Matched Extension. Finally, we evaluated it with 16 participants with motor impairments across four tasks and complemented the study with offline transfer analyses across recognizers and datasets.

We contribute: (1) an accessible interaction logic based on an input-level compatibility layer, which mediates between user expression and recognizer expectations through adversarial assistance rather than user conformity or model retraining; (2) AdaptaFace, a browser-based facial interaction system with user-facing interfaces for gesture registration, assistance tuning, and four personalization modes\footnote{Implementation: \url{https://github.com/SiyuZhang-zsy/AdaptaFace}}; and (3) empirical evidence from formative interviews, a user study, and bounded offline validations showing improved performance, lower workload, greater perceived control, broader usable gesture vocabularies, and the current limits of cross-recognizer transfer.

\section{Related Work}
\label{sec:relatedwork}


\subsection{Facial Gesture Interaction as Accessible Input}
Above-the-neck movements and actions, including head motion, gaze, blinks, eyelid actions, mouth movements, and tongue motion, have long been explored as accessible input for PwM, supporting communication, text entry, cursor control, and environmental interaction~\cite{Betke2002CameraMouse,Bates2003EyeMice,Grauman2003BlinkEyebrow,Majaranta2014EyeHCI,Huo2009TongueWheeled,Kim2013TDS,Fan2020EyelidGestures}. Because these inputs can often be performed without reliable hand use, they provide an important pathway for assistive interaction~\cite{Zhao2022AboveNeck,Sun2024HeadGestures}. Early systems frequently relied on specialized hardware such as eye trackers or instrumented tongue interfaces~\cite{Majaranta2014EyeHCI,Huo2009TongueWheeled,Kim2013TDS}. Commodity cameras and computer-vision toolkits now support real-time webcam and mobile facial analysis without specialist sensing hardware~\cite{Baltrusaitis2016OpenFace,Grishchenko2020AttentionMesh,MediaPipeFaceLandmarkerDocs2024,hu2025openface,Fan2020EyelidGestures}. However, many existing recognizers expose a bounded vocabulary that users must reproduce reliably; recent elicitation studies instead show the value of gestures defined around users' own abilities~\cite{Fan2020EyelidGestures,Zhao2022AboveNeck,Sun2024HeadGestures}. This distinction matters for PwM: feasible movements can be low-amplitude or asymmetric, differ across individuals, and vary with ability and fatigue over time~\cite{Zhang2026BeyondAccuracy,Yamagami2023UpperBodyPersonalization}. Prior work also reports workload, discomfort, and frustration in above-the-neck control, while ability-based and design-justice scholarship cautions against making normative motor assumptions the basis of access~\cite{Bates2003EyeMice,Wobbrock2011ABD,CostanzaChock2020DesignJustice,Zhang2026BeyondAccuracy}. This leaves an unresolved interaction-layer gap: camera access alone does not make a canonical recognizer vocabulary compatible with a person's feasible movement repertoire.

\subsection{Personalization in Recognition-Based Accessibility}
Personalization has become a central strategy for addressing the mismatch between users' abilities and one-size-fits-all interaction systems. In assistive HCI, personalization is important because inclusive systems must adapt to diverse motor repertoires rather than require users to conform to predefined interaction norms~\cite{Wobbrock2011ABD,CostanzaChock2020DesignJustice,Wald2021AIPersonalisation}. Within facial-gesture and broader recognition-based accessibility, personalization spans at least two established strategies. One changes the interaction mapping or interface around a user's abilities without retraining the recognizer. Supple automatically generates interfaces around devices, tasks, preferences, and abilities; MotionBlocks remaps comfortable movements to required virtual-reality input; and gesture-elicitation studies derive vocabularies from movements that users define themselves~\cite{Gajos2010Supple,Wentzel2025MotionBlocks,Zhao2022AboveNeck,Sun2024HeadGestures}. These systems show that personalization can occur through interface generation, remapping, or user-defined vocabularies, with each implementation defining which mappings, parameters, or recognition capabilities are available. A second strategy adapts a recognizer from user-specific data through calibration, fine-tuning, or user demonstrations~\cite{He2019SAGE,Park2019FAZEICCV,Shahi2024SingleDemoCustomization}. Even few-shot methods rely on user-provided calibration samples or demonstrations~\cite{He2019SAGE,Park2019FAZEICCV,Shahi2024SingleDemoCustomization}. When those samples include gaze data, their collection and storage can also raise biometric privacy concerns~\cite{Steil2019PrivacyDP,DavidJohn2022FYEO}. For users whose feasible movements are ability-specific or fatigue-sensitive, providing personal movement data is therefore a consequential design consideration~\cite{Yamagami2023UpperBodyPersonalization,Sun2024HeadGestures}.

Together, these approaches alter either the interaction mapping or the model. A less explored alternative is to improve compatibility at the recognizer input, mediating between user expression and recognizer expectations without changing model parameters.

\subsection{Adversarial Methods as Input-Side Adaptation}

Adversarial perturbations have been widely studied in machine learning as small input modifications that can shift a model's prediction by exploiting its decision boundaries~\cite{Szegedy2013Intriguing,Goodfellow2015ExplainingHarnessing}. Most prior work has framed adversarial examples as threats to robustness and security, motivating extensive research on attacks, defenses, and model stability~\cite{Madry2018PGD,Brown2017AdversarialPatch}. Perturbation-based methods have also been used for interpretability, while adversarial reprogramming has repurposed fixed models for new tasks~\cite{Fong2017MeaningfulPerturbation,Elsayed2018Reprogramming}. These efforts show that controlled input transformations can be used not only to expose vulnerabilities but also to probe or repurpose model behavior.

The accessibility work reviewed above personalizes interaction mappings or models; it does not examine adversarial input transformation as a personalization mechanism. Model-adaptation approaches instead personalize recognition from user-specific calibration samples or demonstrations, even when few-shot methods reduce the amount required~\cite{He2019SAGE,Park2019FAZEICCV,Shahi2024SingleDemoCustomization}. Adversarial reprogramming provides a technical precedent for another possibility: adjusting the input while keeping the target model's parameters fixed~\cite{Elsayed2018Reprogramming}. AdaptaFace applies this input-side logic to accessibility, seeking to make an existing recognizer more compatible with a user's expression patterns. This perspective complements ability-based design, which asks systems to adapt to users' abilities rather than the reverse~\cite{Wobbrock2011ABD}. It also connects to ability-mediating design, which distinguishes adapting a system to an ability from mediating an existing sensorimotor ability so that it can enable a new skill or interaction~\cite{Vatavu2022AbilityMediating}. At the recognizer input, the system preserves the user's performed movement while mediating the image presented to the recognizer. We use this connection as a conceptual lens, not as evidence that the present implementation covers the broader sensorimotor-reality design space.

\section{Formative Study and Design Principles}
\label{sec:formative}

\subsection{Participants and Method}

To ground the design of AdaptaFace, we conducted a formative interview study with 12 PwM (6 women, 6 men; aged 18--36) (Table~\ref{tab:participants}). Participants represented diverse motor conditions and reported varying levels of facial control, asymmetry, fatigue, and movement consistency. We refer to them as FP1--FP12 to distinguish them from evaluation-study participants. We recruited them through social media platforms, non-profit disability organizations, and personal networks. All participants provided informed consent and received a 15 USD honorarium.

Interviews explored three topics: prior experiences with facial and above-neck interaction, perceived barriers in recognizer-defined gesture systems, and expectations for personalization. Remote sessions lasted 45--60 minutes and were recorded with consent. Participants first discussed their abilities and prior interaction experiences, then tried atomic facial actions and reflected on their comfort, feasibility, and distinctness, and finally proposed gestures for example commands using those actions only as a loose reference. We analyzed the interviews using thematic analysis to identify design-relevant patterns related to gesture performance, personalization burden, and desired system behavior. Our aim was not to claim that all reported accessibility barriers were new; rather, the study translated barriers documented in prior work and participant-specific experiences into design principles and system requirements. Further procedural details are provided in Appendix~\ref{sec:formativestudy}.

\begin{table}[t]
\caption{Demographic and functional characteristics of formative-study participants with motor impairments (FP1--FP12).}
\label{tab:participants}
\centering
\footnotesize
\setlength{\tabcolsep}{3pt}
\begin{tabular}{r r l p{2.5cm} l p{2.2cm}}
\toprule
ID & Age & Gender & Health Condition & Since & Functional Traits \\
\midrule
FP1  & 32 & F & SMA II         & 6 mo.  & Ir, Pr, Fa, Dig, Dif \\
FP2  & 31 & M & SCI (C1)       & 10 yrs & Ir, Fa, Dig, Dif, Tre \\
FP3  & 36 & F & SCI (C5)       & 5 yrs  & Ir, Fa, Dig, Dif \\
FP4  & 18 & F & Stroke (hemi.) & 5 yrs  & Ir, Fa, Dig, Dif \\
FP5  & 32 & M & Brainstem Tumor (hemi.) & 3 yrs  & Fa, Dif \\
FP6  & 26 & M & Cerebral Palsy & Birth  & Ir, Fa, Dig, Dif \\
FP7  & 36 & F & SCI (T1--T4)   & 5 yrs  & Ir, Fa, Dig \\
FP8  & 31 & M & Poliomyelitis (para.) & 28 yrs & Pr, Fa, Dig \\
FP9  & 33 & M & Cerebral Palsy & Birth  & Ir, Pr, Dig, Dif, Tre \\
FP10 & 28 & M & Cerebral Palsy & Birth  & Ir, Fa, Dig, Dif, Tre \\
FP11 & 24 & F & Transverse Myelitis & Birth  & Ir, Fa, Dif \\
FP12 & 32 & F & CMT2S          & Birth  & Ir, Fa, Dig, Dif \\
\bottomrule
\end{tabular}

\vspace{2pt}
\footnotesize
\raggedright
\textit{Abbreviations:} SMA = Spinal Muscular Atrophy; SCI = Spinal Cord Injury; CMT = Charcot--Marie--Tooth disease; hemi. = Hemiplegia; para. = Paraplegia. \\
\textit{Functional traits:} Ir = Irreversible; Pr = Progressive; Fa = Rapid fatigue; Dig = Difficulty gross motor; Dif = Difficulty fine motor; Tre = Tremor.
\end{table}

\subsection{Findings}
\label{sec:findings}

\textit{F1. Canonical gestures often failed to fit participants' lived motor repertoires.}
Participants described difficulty reproducing standardized facial gestures due to diverse motor constraints; for instance, FP5 noted, “\emph{At my worst, half my face was numb; it’s hard to make expressions}”, while FP1 explained, “\emph{My muscles are weak; it’s hard for me to lift the corners of my mouth}”. As a result, their comfortable gestures were asymmetric, subtle, and did not necessarily resemble the canonical expressions recognizers were designed for, leading to frequent mismatches in recognition.

\textit{F2. Repeated calibration and retraining imposed disproportionate burden.}
Participants did not reject personalization itself, but many described repeated demonstration, recalibration, and adaptation as tiring and frustrating; as FP4 noted, “\emph{It doesn’t really pick up my expressions unless I exaggerate them, which is tiring}”. This burden was further exacerbated when performance varied across posture, fatigue, or day-to-day condition. For those whose movements differed most from canonical templates, the burden of ``making the recognizer work'' was often highest.

\textit{F3. Participants wanted personalization that preserved their own expressive habits and sense of control.}
Beyond accuracy alone, participants emphasized the importance of using gestures that felt natural, feasible, and socially acceptable to them; as FP11 described, “\emph{If I can nod to confirm, that’s respectful—it’s the same as how I would signal to another person. It feels natural, not like I’m performing something strange for the computer}”. At the same time, participants expressed a need to better understand how their actions were interpreted by the system, suggesting a desire for greater visibility and control rather than opaque automatic adaptation. Together, these accounts suggest that inclusive personalization should improve reliability without removing user agency.

\subsection{Design Principles}

Based on these findings, we derived six design principles for inclusive facial-gesture personalization. Together, these principles point toward a different locus of personalization: a mediating mechanism at the input level that preserves users' own expressive habits while helping existing recognizers interpret them more reliably. Specifically, the system should: (1) support user-compatible gestures rather than only canonical templates; (2) minimize user burden during personalization; (3) enable user-defined gesture vocabularies; (4) preserve user control over personalization; (5) preserve compatibility with existing recognition pipelines; and (6) support low-effort and socially acceptable interaction.

Detailed descriptions of each design principle and how they informed the system design are provided in Appendix~\ref{appendix:extended_formative}.

\section{AdaptaFace: Compatibility Layer}
\label{sec:system}

To instantiate the design principles derived from our formative study, we built AdaptaFace, a real-time facial interaction system organized around a compatibility layer between user expression and recognizer expectations. Rather than changing the downstream recognizer itself, AdaptaFace changes the input pathway into it. This allows users to perform gestures that are feasible and comfortable for them while preserving compatibility with existing recognizers and applications. We describe the compatibility-layer logic, the system pipeline, the adversarial-assistance mechanism, the user controls exposed by the interface, and the four personalization modes supported by the system.

\subsection{Compatibility-Layer Logic}

As explained in the related work, AdaptaFace explores a third strategy of adaptation: \textit{input-side mediation}. A compatibility layer sits between live facial input and the downstream recognizer, transforming the incoming frame so that the recognizer is more likely to interpret a user-compatible gesture as the intended command (Panel B in Figure~\ref{fig:core_mechanism}). The recognizer, its label space, and the application-level command mapping remain unchanged.

We instantiate this compatibility layer through \textit{adversarial assistance}: a bounded, targeted perturbation applied to live facial input to steer recognition toward a registered target output. The goal is not to attack the recognizer in an uncontrolled way, but to mediate between a user's expressive repertoire and the recognizer's predefined decision space while keeping the perturbation visually subtle. This mediation supports two functions. First, it enables \textit{recognition enhancement}, improving reliability for gestures that are already represented in the recognizer's native output space. Second, it enables \textit{gesture vocabulary extension}, allowing a user-defined gesture outside that native vocabulary to be mediated toward an existing recognizer output without retraining the model.

\subsection{System Overview}

AdaptaFace has two phases: registration and runtime interaction. During registration, the user either selects a gesture from a predefined set or records custom gestures and associates them with an intended command. Depending on the personalization mode, the target recognizer output is either predefined or determined through a compatibility-based matching procedure during registration.

Each registered gesture is therefore associated with a target recognizer output before runtime interaction begins. In the fixed modes, this target is specified in advance. In the matched modes, AdaptaFace compares the registered gesture against a set of candidate recognizer outputs during registration and selects the most compatible target once. Runtime interaction does not dynamically switch among multiple targets.

The unit of personalization is therefore a binding among a feasible gesture, a stable internal recognizer target, and an existing application command, rather than a retrained recognizer.

During runtime, AdaptaFace captures live facial frames and checks whether the incoming motion matches one of the registered gesture patterns. When a registered gesture is detected, the compatibility layer computes a bounded perturbation and applies it to the current frame before passing the assisted frame to the downstream recognizer (Panel C in Figure~\ref{fig:core_mechanism}). The assistance is applied toward the target previously assigned to the detected gesture, and the recognizer's output is then mapped to the corresponding application command as in a standard recognition pipeline. AdaptaFace thus personalizes interaction without modifying the recognizer itself or requiring model retraining.

\subsection{Matching, Mediation, and Direct Triggering}
\label{sec:matching_mediation}

Gesture matching and image mediation serve different roles in this pipeline. Matching determines which registered pattern is active and retrieves its internal target. That target is control information within AdaptaFace; it is not sent to the downstream recognizer as a predicted gesture label. Adversarial assistance then uses the target to modify the recognizer-facing image, and only the resulting image is passed downstream.

This distinction provides a recognizer-facing matching-only comparison. If the perturbation is disabled while the matched target remains internal, the downstream recognizer receives the same raw frame as in Baseline; matching alone therefore cannot alter that recognizer's AU output. A different architecture could use the matched label to trigger an application command directly. Such direct triggering is a valid and potentially simpler option when an application exposes a command application programming interface (API) and can adopt AdaptaFace-specific events. It is not the same integration setting evaluated here: it bypasses the image recognizer and changes the application input path. AdaptaFace instead addresses settings in which the downstream image-based recognizer and command mapping remain unchanged.

At runtime, let $\mathcal{R}=\{r_1,\dots,r_M\}$ denote the registered gesture trajectories, each associated with a target recognizer output assigned during registration. Each incoming frame is represented by an unnormalized eight-dimensional AU activation vector $\mathbf{a}_t$, and the matcher maintains the causal window $A_t=\{\mathbf{a}_{t-W+1},\dots,\mathbf{a}_t\}$. For each trajectory $r_j=\{\mathbf{r}_{j,1},\dots,\mathbf{r}_{j,W}\}$, it computes the mean aligned Euclidean distance $D_j(A_t,r_j)=W^{-1}\sum_{i=1}^{W}\lVert\mathbf{a}_{t-W+i}-\mathbf{r}_{j,i}\rVert_2$ and similarity $s_j(A_t,r_j)=1/(1+D_j)$. The highest-similarity template is selected, with ties resolved deterministically by template identifier, and activates only when $s_j\geq\tau_{\mathrm{match}}$. The matcher exposes $T$ as a configurable stability requirement; the study configuration used $W=1$, $\tau_{\mathrm{match}}=2/3$ (equivalently, Euclidean distance at most $0.5$), and $T=3$, so the same template had to remain top-ranked and above threshold for three consecutive frame-level windows. Once confirmed, the detected trajectory retrieves its previously assigned internal target and conditions frame-level perturbation generation; the matching label itself is not passed to the downstream recognizer. Appendix~\ref{app:runtime_detection} provides the expanded onset-detection description.

\subsection{Adversarial Assistance Mechanism}
\label{sec:core_mechanism}

To generate assistance efficiently, AdaptaFace uses a differentiable proxy model to estimate how small pixel-space changes affect recognizer outputs. We instantiate this process using a projected gradient descent (PGD)-style targeted optimization procedure with a random start. Given an input frame $x \in \mathbb{R}^{H \times W \times C}$ and a target recognizer output $y^*$ associated with the detected registered gesture, the system iteratively computes a bounded perturbation $\delta$ that reduces the proxy-model loss for $y^*$ while constraining $\|\delta\|_\infty \le \epsilon$. We use a small number of iterations in practice to preserve real-time performance and keep the perturbation visually subtle. Formally, the perturbation is computed as
\begin{equation}
\delta^* = \arg\min_{\|\delta\|_\infty \le \epsilon} \mathcal{L}\big(f_{\text{proxy}}(x+\delta), y^*\big),
\end{equation}
where $f_{\text{proxy}}(\cdot)$ is a differentiable proxy model, $\mathcal{L}$ is a targeted loss, and $\epsilon$ bounds perturbation magnitude.

The proxy model is used to generate perturbations; the downstream recognizer's parameters and output vocabulary remain unchanged. During registration, AdaptaFace determines which recognizer output should serve as the target for each registered gesture; during runtime, it applies frame-level assistance toward that assigned target whenever the corresponding gesture is detected. The perturbation is nevertheless proxy- and output-space-dependent. We instantiate the mechanism with differentiable facial-AU models and evaluate cross-recognizer transfer separately in Section~\ref{sec:offline_validation}. Additional implementation details are provided in Appendix~\ref{app:PGD}.

\subsection{Four Personalization Modes}

AdaptaFace defines four personalization modes from two axes: whether interaction stays within or extends beyond the recognizer vocabulary (\textit{enhancement} vs.\ \textit{extension}) and whether the assistance target is predefined or selected through compatibility-based matching during registration (\textit{fixed} vs.\ \textit{matched}). In fixed modes, AdaptaFace steers recognition toward a predefined target recognizer output. In matched modes, AdaptaFace compares the user's registered gesture against a set of candidate recognizer outputs during registration and selects the most compatible target once. During runtime, assistance is applied to the target assigned during registration. In both cases, the downstream recognizer is unchanged, and the compatibility layer adapts the input pathway that leads into it. Together, the four modes span fixed versus matched target assignment and enhancement versus extension.

\paragraph{Fixed Enhancement.}
Users interact through a predefined gesture set specified by the interface, with fixed gesture--command mappings. AdaptaFace applies assistance toward a fixed recognizer-supported target, improving reliability for gestures that already exist within the interface vocabulary. This mode is closest to a conventional recognizer-centric pipeline, in which neither the gesture vocabulary nor the mapping is personalized.

\paragraph{Matched Enhancement.}
This mode is designed for gesture-selectable interfaces. During registration, AdaptaFace compares the user's performed gesture against the set of gestures supported by the interface and selects the candidate gesture that is most compatible. The selected match is then used as the target recognizer output during runtime interaction. This mode allows personalization within a predefined vocabulary by identifying the interface-supported gesture that best aligns with the user's performance, while keeping the available gesture set bounded by the designer.

\paragraph{Fixed Extension.}
Users define and record their own gesture--command bindings. Each custom gesture is associated with a fixed target recognizer output selected during registration. This mode extends interaction beyond a closed predefined vocabulary while keeping the assistance target stable during runtime. It supports user-defined input without requiring changes to the downstream recognizer.

\paragraph{Matched Extension.}
Users define and record their own gestures, but instead of manually binding each gesture to a fixed target, the system compares the registered gesture with candidate recognizer outputs during registration and selects the most compatible target. This mode combines user-defined input with system-selected target matching and runtime input-side mediation toward that selected target. It is intended to support subtle, asymmetric, or only partially aligned gestures and enables gesture vocabulary extension.

\subsection{Parameter Settings}

The compatibility layer is governed by a small set of optimization parameters. In the study, these were exposed as lightweight user-facing controls; in the offline validations, they were fixed for consistency ($\epsilon=8/255$, $\alpha=3/255$, $K=3$). These controls allowed participants to balance steering effectiveness, responsiveness, stability, and visual subtlety without requiring developer intervention.

Table~\ref{tab:user-controls} shows the user-facing controls and intended interaction effects. Higher strength, precision, and iteration settings increase the likelihood of successfully steering recognition toward the intended target, but may also introduce stronger visual effects, less smooth updates, or higher latency. The recognition-speed preset and gesture cooldown let users trade off responsiveness against stability and accidental repeated triggers. Internally, assistance strength, assistance precision, and iteration steps correspond to the perturbation bound $\epsilon$, update step size $\alpha$, and number of optimization steps $K$, respectively, but are exposed to participants through mapped interface ranges rather than raw optimization values. Figure~\ref{fig:evaluation-interfaces} shows how these controls are surfaced in the interface.

\begin{table*}[t]
\centering
\small
\setlength{\tabcolsep}{5pt}
\renewcommand{\arraystretch}{1.08}
\caption{User-facing controls in the participant study. Values refer to interface-level controls rather than raw optimization parameters. In the offline validations, perturbation settings were fixed to $\epsilon=8/255$, $\alpha=3/255$, and $K=3$.}
\label{tab:user-controls}
\begin{tabular}{p{3.5cm} p{3.6cm} p{2.4cm} p{5.8cm}}
\toprule
\textbf{Control} & \textbf{Role} & \textbf{Range / Default} & \textbf{Effect} \\
\midrule
Assistance strength ($\epsilon$)
& Perturbation budget
& 1--20 / 5
& steering $\uparrow$; too high $\rightarrow$ overshoot \\

Assistance precision ($\alpha$)
& Update step size
& 1--10 / 5
& aggressiveness $\uparrow$; too high $\rightarrow$ less smooth \\

Iteration steps ($K$)
& Refinement steps
& 1--10 / 3
& accuracy $\uparrow$, latency $\uparrow$ \\

Recognition speed preset
& Response--stability balance
& 1--5 / 3
& faster response $\leftrightarrow$ calmer behavior \\

Gesture cooldown (ms)
& Trigger interval
& 100--2000 / 400
& repeats $\downarrow$; rapid actions slower \\
\bottomrule
\end{tabular}
\end{table*}

\subsection{Implementation and Deployment}

We implemented AdaptaFace as a modular framework with a Python-based core that separates live input capture, compatibility-layer processing, and downstream recognition. In the current implementation, the compatibility layer is instantiated with differentiable proxy models based on OpenFace~3.0 and OpenGraphAU~\cite{Luo2022MEGraphAU}. In the online study configuration, the same OpenFace-based AU model supplied both the perturbation-generation objective and the post-assistance AU scores; the online evaluation therefore tested an aligned proxy/evaluator setting rather than an independent downstream recognizer. Participants accessed a browser-based client that supported gesture registration, assistance tuning, and real-time interaction. The study interface was exposed remotely via ngrok, while all computation ran on a local laptop equipped with an NVIDIA RTX 3060 graphics processing unit (GPU). Under this setup, the system maintained end-to-end latency below 80\,ms per frame, supporting real-time assistance in the study setting. The same hardware configuration was also used for local testing. This deployment demonstrates an insertion point immediately before an AU-based recognizer; support for other recognizer families depends on a compatible proxy and target space and is not established by the present implementation.

\section{Evaluation with participants with motor impairments}
\label{sec:user_study}
\begin{figure*}[t]
  \centering
  \includegraphics[width=\textwidth]{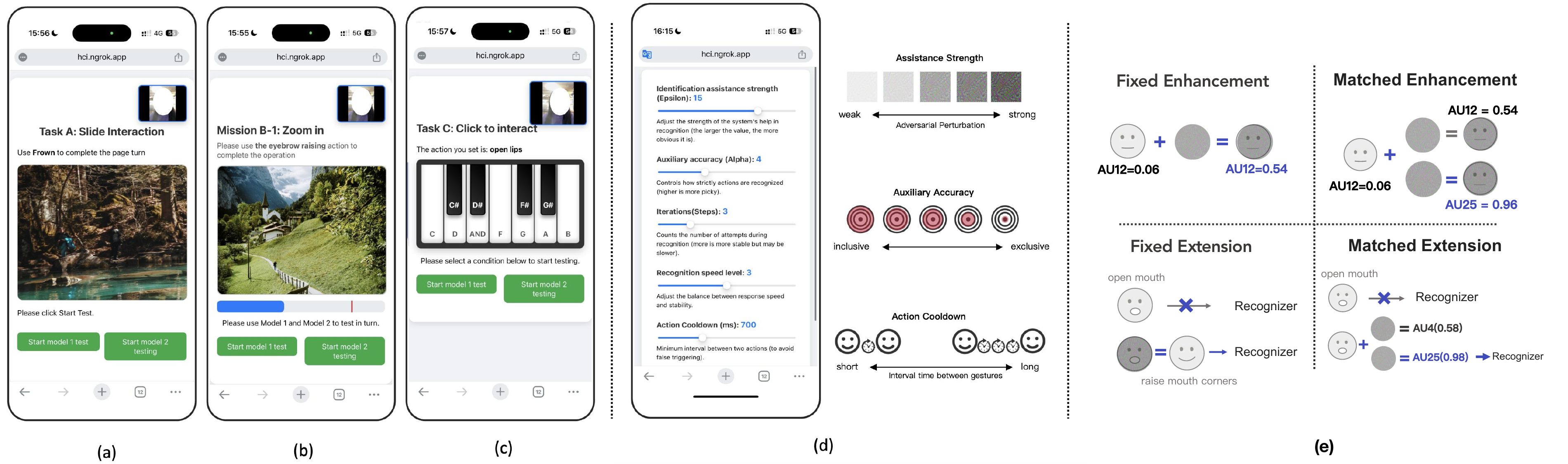}
\caption{Evaluation interfaces and modes: (a) Swipe, (b) Zoom, (c) Click, (d) assistance controls, and (e) Fixed Enhancement, Matched Enhancement, Fixed Extension, and Matched Extension.}
  \Description{Screenshots show swipe, zoom, and click tasks, five assistance controls, and a two-by-two mode design. The modes cross fixed versus matched target selection with enhancement versus extension of the recognizer vocabulary.}
  \label{fig:evaluation-interfaces}
\end{figure*}

We evaluated AdaptaFace with participants with motor impairments to assess how well input-level adversarial assistance supports facial-gesture interaction across the four personalization modes. Our goals were: (G1) to quantify the performance benefits of input-level assistance across tasks and gesture types, (G2) to evaluate usability, cognitive effort, and perceived control during interaction, and (G3) to understand user preferences, perceived mode suitability, and how different modes aligned with individual abilities.

\subsection{Study Design}
\label{sec:design}
We used a within-subjects design with three independent variables: \textit{Technology} (Baseline vs.\ AdaptaFace), \textit{Task} (Swipe, Click, Zoom In, Zoom Out), and \textit{Personalization Mode} (Fixed Enhancement, Matched Enhancement, Fixed Extension, and Matched Extension). This yielded a $2 \times 4 \times 4$ factorial design. Each Technology $\times$ Task $\times$ Mode task block contained five action repetitions, resulting in 32 planned task blocks and 160 planned action repetitions per participant, or 512 task blocks and 2,560 action repetitions overall. We counterbalanced the sequence of technology conditions across participants, balanced task order using a Latin square, and randomized mode order within each Technology $\times$ Task block. The design compares the complete Baseline and AdaptaFace pipelines across personalization modes; it does not independently ablate matching, target selection, and perturbation generation.

\subsection{Participants}

\begin{table}[t]
\centering
\caption{Participant characteristics for the evaluation study ($N=16$). FS identifies participants who also took part in the formative study.}
\label{tab:participants2}
\small
\setlength{\tabcolsep}{6pt}

\begin{tabular}{r r l l l l l}
\toprule
ID & Age & Gender & Condition & Device & Wheel. & FS \\
\midrule
EP1  & 32 & Female & SMA2 & Phone & Yes & FP1 \\
EP2  & 36 & Female & SCI(C5) & Phone & Yes & FP3 \\
EP3  & 37 & Male   & DMD & Phone & Yes & -- \\
EP4  & 19 & Female & Stroke & Phone & No & FP4 \\
EP5  & 28 & Female & SMA2 & Phone & Yes & -- \\
EP6  & 32 & Female & CMT2S & Phone & Yes & FP12 \\
EP7  & 24 & Female & SMA2 & Computer & Yes & -- \\
EP8  & 40 & Female & Paralysis & Phone & Yes & -- \\
EP9  & 27 & Female & SMA3 & Phone & Yes & -- \\
EP10 & 28 & Female & SMA & Computer & Yes & -- \\
EP11 & 25 & Female & SCI & Phone & Yes & -- \\
EP12 & 26 & Male   & CP & Phone & No & FP6 \\
EP13 & 27 & Male   & SCI(T3) & Phone & Yes & -- \\
EP14 & 28 & Male   & SMA3 & Phone & Yes & -- \\
EP15 & 35 & Female & SB & Phone & Yes & -- \\
EP16 & 30 & Male   & BD & Phone & Yes & -- \\
\bottomrule
\end{tabular}
\vspace{2pt}
\raggedright
\footnotesize
\textit{Abbreviations:} SMA = Spinal Muscular Atrophy; SCI = Spinal Cord Injury; DMD = Duchenne Muscular Dystrophy; CMT = Charcot--Marie--Tooth disease; CP = Cerebral Palsy; SB = Spina Bifida; BD = Behçet’s Disease; Wheel. = Wheelchair; FS = Formative Study. \\
\end{table}

We recruited 16 adults with motor impairments that limited reliable hand use but preserved some above-neck control (\(N{=}16\); 11 female, 5 male; age 19--40, mean \(M{=}28.9\), standard deviation \(SD{=}5.1\)) (Table~\ref{tab:participants2}). Participants were recruited via disability organizations, online support communities, and word of mouth. Inclusion required (i) reduced or unreliable upper-limb function for everyday device use and (ii) ability to perform at least one facial or head gesture. Participants represented diverse impairment etiologies, including cerebral palsy, spinal cord injury, stroke, spinal muscular atrophy, Duchenne muscular dystrophy, Charcot--Marie--Tooth disease, spina bifida, Behçet’s disease, and other paralysis. 14 participants used wheelchairs, and all reported prior experience with above-neck input such as gaze, head movement, or facial actions. 5 participants had also taken part in the formative study. All sessions were conducted remotely to remove travel barriers. The study allowed breaks, and supported alternatives such as typing. Both the formative study and the subsequent evaluation were approved by the ethics committee of the University of Bristol. All participants provided informed consent, and each evaluation participant received a 15 USD honorarium.

\subsection{Apparatus, Tasks, and Procedure}
\label{sec:tasks_procedure}

Participants completed the remote study using their everyday phones or laptops and built-in webcams, connecting to AdaptaFace through a secure browser client from their own environments and preferred postures. The system was a zero-install web interface using webcam input to provide real-time input-side assistance and adapted feedback. We logged lightweight metadata required for analysis, including timestamps, task events, recognition outcomes, confidence scores, and gesture--command mappings.

We evaluated AdaptaFace on four interaction tasks: \textit{Swipe} (continuous navigation), \textit{Click} (discrete selection), \textit{Zoom In}, and \textit{Zoom Out} (bidirectional scaling) (Figure~\ref{fig:evaluation-interfaces}). These tasks capture different interaction demands for repetition, precision, and continuous control.

Each session lasted 30--60 minutes depending on pace and breaks. After consent and setup, participants joined a video call, received an introduction to the task interfaces, and completed a practice round. Participants then completed all experimental blocks. In modes other than \textit{Fixed Enhancement}, they could select or define gestures that suited their abilities; in matched modes, target recognizer outputs were assigned automatically during registration. At the start of each block, participants could review or re-record gestures if needed. After each block, they rated ease of use and perceived control. After the session, they completed the System Usability Scale (SUS) and the NASA Task Load Index with raw scoring (NASA Raw TLX) separately for Baseline and AdaptaFace, followed by a short semi-structured exit interview on gesture preferences, perceived control, workload, and social acceptability. To reduce fatigue, participants were encouraged to take breaks between tasks and could pause at any time or split the session across two sittings if needed.

\subsection{Measures}
\label{sec:measures}

We collected objective performance metrics, subjective experience measures, and records of gesture creation and mappings in personalization modes.

\textit{Objective performance.}
For each task block, we logged task completion time (TCT), completion rate (CR), success-normalized task time (SNT), and event-level recognition confidence. TCT measured the elapsed time required to complete the assigned interaction sequence. CR was the proportion of required action repetitions successfully completed within the block. For inferential analysis, we additionally derived full-block completion, a binary indicator of whether all required repetitions in a block were completed. SNT normalized task time by the number of successful repetitions, enabling comparison when completion was partial. Recognition confidence was computed at the event level rather than per frame.

\textit{Subjective experience.}
After each block, participants provided two 5-point Likert ratings on ease of use and perceived control. After completing all tasks, they filled out the System Usability Scale (SUS) and the NASA Raw TLX separately for Baseline and AdaptaFace.

\textit{Gesture creation and mappings.}
In personalization modes, we recorded the chosen gesture type, the bound command, whether the gesture extended beyond the recognizer’s exposed AU vocabulary, and the associated parameter settings. This enabled us to analyze gesture diversity, vocabulary extension, and how participants tuned assistance.

\textit{Qualitative feedback.}
A semi-structured exit interview gathered reflections on gesture preferences, perceived strengths and weaknesses of AdaptaFace, trust and feedback, social acceptability, and long-term use.

\begin{table}
\centering
\small
\setlength{\tabcolsep}{5pt}
\caption{Performance by technology. $\Delta$ shows AdaptaFace minus Baseline (units follow each column); Baseline is the unmodified OpenFace-3.0 recognizer with default AU thresholds.}
\begin{tabular}{lcccc}
\toprule
 & CR\,\% (SD) & TCT\,s (SD) & SNT\,s (SD) & Conf. (SD) \\
\midrule
Baseline    & 59.7 (44.7) & 16.6 (16.7) & 37.7 (151.3) & 0.65 (0.43) \\
AdaptaFace  & 90.5 (25.6) & 13.4 (9.0)  & 15.9 (16.1)  & 0.89 (0.28) \\
$\Delta$ (A$-$B) & +30.8 & \,-3.2 & \,-21.8 & +0.24 \\
\bottomrule
\end{tabular}

\label{tab:overall_perf}
\end{table}

\begin{table*}[t]
\centering
\caption{Performance by mode (mean (SD)).}
\small
\setlength{\tabcolsep}{5pt}
\begin{tabular}{ccccccccc}
\toprule
Mode & \multicolumn{2}{c}{CR\,\%} & \multicolumn{2}{c}{TCT\,s} & \multicolumn{2}{c}{SNT\,s} & \multicolumn{2}{c}{Conf.} \\
\cmidrule(lr){2-3} \cmidrule(lr){4-5} \cmidrule(lr){6-7} \cmidrule(lr){8-9}
 & Baseline & AdaptaFace & Baseline & AdaptaFace & Baseline & AdaptaFace & Baseline & AdaptaFace \\
\midrule
Fixed Extension     & 27.9 (40.8) & 82.4 (33.0) & 22.0 (26.6) & 16.6 (10.7) & 135.4 (380.8) & 23.5 (26.9) & 0.38 (0.45) & 0.82 (0.36) \\
Matched Extension   & 55.8 (45.4) & 91.9 (24.1) & 14.9 (10.2) & 11.7 (8.7)  & 30.9 (48.3)  & 13.4 (10.4)  & 0.65 (0.43) & 0.91 (0.24) \\
Fixed Enhancement   & 77.3 (35.2) & 94.8 (17.2) & 15.6 (12.3) & 12.6 (7.8)  & 17.7 (17.9)  & 13.9 (9.1)   & 0.83 (0.31) & 0.92 (0.21) \\
Matched Enhancement & 77.6 (37.7) & 92.7 (24.5) & 13.9 (11.4) & 12.5 (7.7)  & 16.2 (14.8)  & 13.3 (8.9)   & 0.76 (0.35) & 0.89 (0.26) \\
\bottomrule
\end{tabular}

\label{tab:perf_by_mode}
\end{table*}

\begin{table*}[t]
\centering
\small
\caption{System-level scales. Higher is better for SUS; lower is better for NASA Raw TLX.}
\label{tab:subjective-all-in-one-row}

\begin{subtable}[t]{0.24\linewidth}
\centering
\caption{SUS (0--100)}
\setlength{\tabcolsep}{2pt}
\begin{tabular}{lrrr}
\toprule
System & Mean & SD & Median \\
\midrule
Baseline   & 47.3 & 15.7 & 50.0 \\
AdaptaFace & 70.8 & 14.7 & 72.5 \\
\bottomrule
\end{tabular}
\end{subtable}\hfill
\begin{subtable}[t]{0.24\linewidth}
\centering
\caption{NASA Raw TLX}
\setlength{\tabcolsep}{2pt}
\begin{tabular}{lrrr}
\toprule
System & Mean & SD & Median \\
\midrule
Baseline   & 5.61 & 1.54 & 5.83 \\
AdaptaFace & 4.21 & 1.56 & 4.50 \\
\bottomrule
\end{tabular}
\end{subtable}\hfill
\begin{subtable}[t]{0.48\linewidth}
\centering
\caption{NASA Raw TLX subscale deltas (AdaptaFace $-$ Baseline)}
\setlength{\tabcolsep}{2pt}
\begin{tabular}{lrrrrrr}
\toprule
Subscale & Cognitive & Physical & Temporal & Effort & Frustration & Performance \\
\midrule
$\Delta$  & $-0.87$ & $-1.33$ & $-1.27$ & $-1.00$ & $-2.00$ & $+1.93$ \\
\bottomrule
\end{tabular}
\end{subtable}
\end{table*}

\subsection{Data Analysis}
\label{sec:data_analysis}

We analyzed binary full-block completion using a binomial generalized estimating equation (GEE) with a logit link, and continuous repeated-measures outcomes using linear mixed-effects models (LMMs) with participant-level random intercepts. Task completion time (TCT), success-normalized task time (SNT), ease, and perceived control were analyzed within this repeated-measures framework, while CR percentages were summarized descriptively. SUS and NASA Raw TLX, which were collected per technology, were compared descriptively at the participant level between Baseline and AdaptaFace. No Technology $\times$ Task or Technology $\times$ Mode interaction reached significance; we therefore focus on Technology main effects, while task- and mode-level breakdowns are interpreted descriptively. Model equations and fixed-effect tables are reported in Appendix Section~\ref{app:stats}.

Qualitative interview responses were analyzed using thematic analysis. Two researchers reviewed transcripts and notes iteratively, grouped recurring observations into themes, and used these themes to interpret participants’ gesture preferences, perceived control, workload, and social acceptability. In the main paper, we report qualitative findings primarily to contextualize the quantitative results rather than as a standalone qualitative contribution.

\subsection{Results}
\label{sec:userstudy_results}

Results follow G1--G3. We first report objective performance and mode-level patterns, then subjective experience, and finally gesture-vocabulary extension. Qualitative accounts then examine how participants experienced the personalization choices underlying those outcomes.

\subsubsection{Overall Objective Performance}
\label{sec:results_overall}

Across tasks and personalization modes, AdaptaFace improved all objective measures (Table~\ref{tab:overall_perf}). Mean within-block CR increased from \textbf{59.7\%} to \textbf{90.5\%}; a GEE on binary full-block completion showed a significant Technology effect ($\beta{=}1.12$, Wald $z{=}11.86$, $p{<}.001$). TCT decreased from \textbf{16.6\,s} to \textbf{13.4\,s} (LMM: $\beta{=}-3.18$, $z{=}-7.92$, $p{<}.001$, 95\% confidence interval (CI) [$-3.97$, $-2.40$]); SNT decreased from \textbf{37.7\,s} to \textbf{15.9\,s} ($\beta{=}-12.16$, $z{=}-12.63$, $p{<}.001$, 95\% CI [$-14.05$, $-10.27$]). Mean event-level confidence increased from \textbf{0.65} to \textbf{0.89}.

Completion improved in every mode (Table~\ref{tab:perf_by_mode}): \textit{Fixed Enhancement}, \textbf{77.3\%--94.8\%}; \textit{Matched Enhancement}, \textbf{77.6\%--92.7\%}; \textit{Fixed Extension}, \textbf{27.9\%--82.4\%}; and \textit{Matched Extension}, \textbf{55.8\%--91.9\%}. The enhancement modes improved reliability while retaining the recognizer's existing vocabulary, whereas the largest descriptive changes appeared in the extension modes, where personalized gestures were unlikely to succeed without assistance. SNT showed the same pattern. Because no Technology $\times$ Task or Technology $\times$ Mode interaction was significant, these breakdowns are descriptive rather than confirmed effect-size differences.

Completion and confidence improved across all four tasks, whereas time measures did not improve uniformly (Appendix Table~\ref{tab:perf_by_task}). \textit{Zoom In} was the most variable Baseline task, while \textit{Click} had the highest completion and lowest AdaptaFace task time. Together, the mode and task patterns show two complementary outcomes: more reliable use of recognizer-supported gestures and practical use of personalized gestures outside the exposed vocabulary.

Runtime template matching, used in all four AdaptaFace modes and distinct from registration-time target matching in the Matched modes, was activated at least once in \textbf{95.6\%} of the analyzed AdaptaFace task blocks. Among successful runtime matching events, Euclidean matching distance had a median of \textbf{0.1067} (interquartile range (IQR) [0.0256, 0.2683]), a mean of \textbf{0.1549} (SD = 0.1469), and a range of [\textbf{0.0000}, \textbf{0.4998}].

\subsubsection{Subjective Experience}
\label{sec:results_subjective}

AdaptaFace increased ease from \textbf{3.70} to \textbf{4.35} ($\beta{=}0.67$, $z{=}13.12$, $p{<}.001$, 95\% CI [$0.57$, $0.78$]) and perceived control from \textbf{3.15} to \textbf{4.27} ($\beta{=}1.12$, $z{=}20.48$, $p{<}.001$, 95\% CI [$1.02$, $1.23$]). SUS increased from \textbf{47.3} to \textbf{70.8}, and NASA Raw TLX decreased from \textbf{5.61} to \textbf{4.21} (Table~\ref{tab:subjective-all-in-one-row}). Workload reductions were broad rather than isolated to one dimension: cognitive, physical, temporal, effort, and frustration scores decreased, while perceived performance improved. These converging measures indicate that participants experienced the assisted pipeline as both easier to operate and more controllable. Appendix~\ref{app:extended_userstudy} reports task- and mode-level ratings.

\subsubsection{Gesture Vocabulary Extension}
\label{sec:results_gesture_creation}

All \textbf{16 participants} introduced personalized gestures in extension modes. The resulting repertoire included oral/lip actions such as \textit{Pout}, \textit{Lip Press}, and \textit{Mouth Widen}; ocular/gaze actions such as \textit{Close Right Eye}, \textit{Eye Look Up}, and \textit{Half-squinted Eyes}; and cheek- and tongue-based gestures. Oral/lip gestures were most frequent (\textbf{40} instances), followed by ocular/gaze/brow (\textbf{38}), cheek (\textbf{5}), and tongue (\textbf{4}).

Several gestures, including \textit{Pout}, \textit{Lip Press}, and \textit{Close Right Eye}, were reused across tasks. Participants were therefore developing small personal repertoires rather than selecting isolated substitutions for individual tasks. This distinguishes extension from enhancement: the enhancement modes improved reliability within the recognizer's vocabulary, while the extension modes broadened which movements could function as input at all. Appendix~\ref{app:extended_userstudy} reports the full inventory.

\subsection{Qualitative Feedback}
\label{sec:results_qualitative}

Interviews showed that AdaptaFace changed not only how reliably gestures were recognized, but also which movements participants felt they could comfortably and confidently use as input. We organize these accounts into three themes connecting performance to the experience and tradeoffs of personalization.

\subsubsection{AdaptaFace preserved participants’ own gesture abilities and preferences}

Participants valued gestures aligned with their motor abilities rather than canonical expressions. Common preset gestures could be physically demanding or unnatural; one participant noted that \textit{“for many people with severe motor impairments, opening the mouth wide is very tiring.”} EP1 explained the benefit of extension more directly: \textit{“Even if I can’t open my mouth wide, I can still use a small version in custom mode,”} while EP6 noted, \textit{“Everyone has different strengths; customization lets you pick what you can actually do.”} Rather than treating deviation from a canonical form as failure, personalization preserved movements that were sustainable and personally appropriate.

\subsubsection{AdaptaFace reduced adaptation burden and improved perceived ease of use}

Participants contrasted AdaptaFace with Baseline in terms of both physical and cognitive effort. Baseline often required exaggerated expressions (EP3: \textit{“You have to open your mouth really wide”}), whereas EP4 described AdaptaFace as \textit{“less tiring.”} EP13 highlighted a different burden: \textit{“If you set it up for me, I might not remember\ldots My own gestures are much easier to remember.”} Self-defined mappings therefore reduced the need to repeatedly exaggerate movements and the need to remember system-imposed gestures. This helps contextualize the improvements in ease, control, and workload as consequences of a better fit between interaction demands and participants' existing habits.

Perceived control did not arise from automation alone. Mirror views, live labels, and immediate responses helped participants judge whether a gesture had been recognized, understand the current gesture--command mapping, and decide whether to re-record a gesture or adjust assistance. This feedback made personalization legible rather than invisible. At the same time, some controls and parameter names felt technical, creating a tension between configurability and setup burden. Adjustable assistance supported agency when participants could see its effects, but additional control was not automatically useful unless the interface made those controls understandable.

\subsubsection{AdaptaFace expanded participants’ usable gesture vocabulary}

Participants treated extension as more than a one-time substitution. EP8 said, \textit{“I like to switch expressions when I get tired of one,”} while EP13 contrasted selectable mode's eight choices with custom mode, where \textit{“I can just do whatever I want.”} These accounts reveal a tradeoff across modes: selectable gestures offered a bounded set that could reduce setup decisions, while custom gestures improved personal fit and memorability. Participants valued small repertoires that could accommodate changes in fatigue, task, or context rather than one permanently optimal gesture. Preferences for subtle gestures in public also linked vocabulary extension to social comfort. Together, these findings suggest that accessible personalization should support families of alternatives and allow mappings to evolve with the user's situation.



\section{Offline Validation: Cross-Dataset, Cross-Model, and Model-Transfer Analyses}
\label{sec:offline_validation}
\begin{figure*}[t]
  \centering
  \caption{Examples from the three offline validation settings: cross-dataset (left), cross-model (middle), and model transfer (right). Each triplet shows the original input, the assisted image, and a normalized visualization of the perturbation. Numbers below the original and assisted images indicate target confidence before and after assistance.}
  \includegraphics[width=\textwidth]{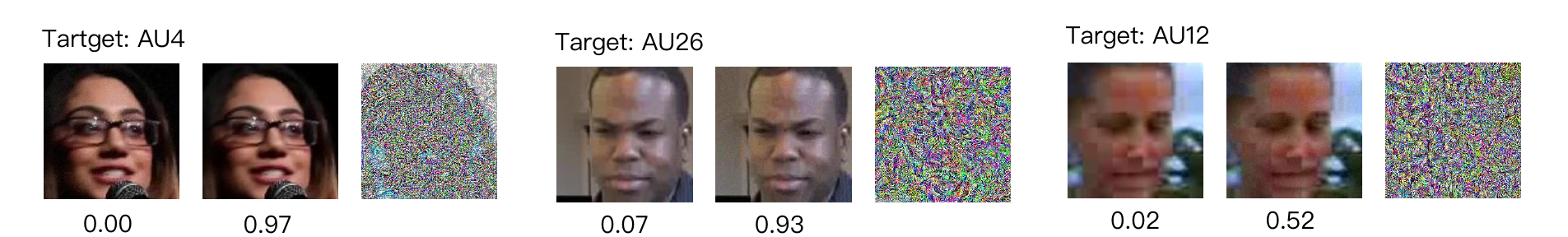}
  \Description{Three groups compare original facial images, assisted images, and magnified perturbation visualizations. The groups illustrate cross-dataset, cross-model, and cross-recognizer-transfer settings, with target-confidence values shown before and after assistance.}
  \label{fig:validation}
\end{figure*}

\begin{table*}[t]
\centering
\caption{Aggregate offline validation results across three settings: cross-dataset, cross-model, and model-transfer. S $ =$ substitution (target$ = 0$) tests whether assistance can steer recognition toward a target AU that is initially absent, whereas E $ =$ enhancement (target$ = 1$)  tests whether assistance can strengthen recognition of a target AU that should become active. Conf.\ reports target-confidence success rate, and Top-1 reports the proportion of examples for which the target AU became the top-ranked prediction. Gains are computed relative to the baseline condition.}
\label{tab:validation_main_split}
\small
\setlength{\tabcolsep}{4.5pt}
\begin{tabular}{llcccccc}
\toprule
Setting & Case & Base Conf. & Adv Conf. & Conf. Gain & Base Top-1 & Adv Top-1 & Top-1 Gain \\
\midrule
Cross-dataset & Substitution  & 0.0223 & 0.5248 & 0.5026 & 0.0401 & 0.5917 & 0.5516 \\
Cross-dataset & Enhancement   & 0.3238 & 0.7591 & 0.4352 & 0.3853 & 0.7863 & 0.4011 \\

Cross-model   & Substitution  & 0.0087 & 0.4378 & 0.4291 & 0.0174 & 0.4387 & 0.4212 \\
Cross-model   & Enhancement   & 0.3692 & 0.7175 & 0.3483 & 0.4553 & 0.7179 & 0.2626 \\

Model-transfer & Substitution  & 0.0368 & 0.0432 & 0.0064 & 0.0584 & 0.0812 & 0.0228 \\
Model-transfer & Enhancement  & 0.2407 & 0.2665 & 0.0258 & 0.3022 & 0.3475 & 0.0452 \\
\bottomrule
\end{tabular}
\end{table*}

We next examine whether the compatibility-layer mechanism shows signs of transfer beyond the exact user-study deployment setting. We conducted offline validation on Aff-Wild2, a large in-the-wild facial-behavior dataset with AU annotations \cite{zafeiriou2017aff,kollias2019deep,kollias2019expression,kollias2023abaw2}. This dataset provides diverse, unconstrained facial behavior at scale while covering the shared AUs used in our system, making it suitable for controlled validation of the underlying assistance mechanism. 

The three settings progressively relax alignment between perturbation generation and evaluation. In \textit{cross-dataset validation}, perturbations were generated with OpenFace and evaluated by OpenFace on Aff-Wild2. In \textit{cross-model validation}, the same validation logic was instantiated with OpenGraphAU, i.e., perturbations were generated and evaluated within a different recognizer family on the same dataset. In \textit{model-transfer validation}, perturbations generated with OpenGraphAU were evaluated by OpenFace, testing whether the input-side effect transferred across recognizers.

Following the logic of AdaptaFace, we focus on two state types. \textit{Enhancement} tests whether assistance can strengthen recognition of a target AU that should become active. \textit{Substitution} tests whether assistance can steer recognition away from a present source AU toward an absent target AU. For each setting, we compared Baseline, adversarial assistance, and random perturbations. Table~\ref{tab:validation_main_split} reports the aggregate baseline and adversarial-assistance results, while the random-perturbation condition remained close to baseline across settings and is summarized in the text below.

Across both state types, cross-dataset validation showed the strongest aggregate evidence of successful steering. In the enhancement case, target-confidence success increased from \textbf{0.3238} to \textbf{0.7591}, and target top-1 from \textbf{0.3853} to \textbf{0.7863}. In the substitution case, target-confidence success increased from \textbf{0.0223} to \textbf{0.5248}, and target top-1 from \textbf{0.0401} to \textbf{0.5917}. These gains suggest that the compatibility-layer mechanism can remain effective under dataset shift when perturbation generation and evaluation are still aligned within the same recognizer.

Cross-model validation with OpenGraphAU showed the same overall direction, indicating that the mechanism is not tied to OpenFace alone. In enhancement, target-confidence success increased from \textbf{0.3692} to \textbf{0.7175}, and target top-1 from \textbf{0.4553} to \textbf{0.7179}. In substitution, target-confidence success increased from \textbf{0.0087} to \textbf{0.4378}, and target top-1 from \textbf{0.0174} to \textbf{0.4387}. At the same time, the pair-level results were more target-dependent than in the cross-dataset setting: some AU pairs approached near-saturation under assistance, while others remained comparatively difficult.

By contrast, model-transfer validation from OpenGraphAU to OpenFace was weaker. In enhancement, target-confidence success increased only from \textbf{0.2407} to \textbf{0.2665}, and target top-1 from \textbf{0.3022} to \textbf{0.3475}. In substitution, target-confidence success increased from \textbf{0.0368} to \textbf{0.0432}, and target top-1 from \textbf{0.0584} to \textbf{0.0812}. These modest aggregate gains suggest that transfer across recognizers is possible in some cases, but remains less reliable than settings in which perturbation generation and evaluation are aligned.

Across all settings, random perturbations remained close to baseline and consistently below adversarial assistance. This indicates that the observed gains were not explained by perturbation magnitude alone, but depended on targeted steering toward the intended AU.

These results clarify the evaluated boundary of AdaptaFace. Input-level assistance transfers most strongly when the perturbation generator and evaluator are aligned, remains viable when instantiated with a different recognizer family, and becomes substantially less reliable under cross-recognizer transfer.

\section{Discussion}
\label{sec:discussion}

\subsection{Personalization as Accessibility Mediation}

AdaptaFace reframes adversarial perturbation from a robustness threat into an accessibility mechanism. Rather than asking users to make their movements conform to a canonical vocabulary, it mediates the recognizer-facing image so that feasible movements can become usable commands. This complements ability-based design, which asks systems to adapt to users' abilities~\cite{Wobbrock2011ABD}, and ability-mediating design, which foregrounds how existing sensorimotor abilities can be mediated to enable new interaction abilities~\cite{Vatavu2022AbilityMediating}.

Personalization is therefore the central contribution. Enhancement modes improve use of the recognizer's existing vocabulary, while extension modes let participants use subtler, more memorable, or personally meaningful movements without retraining the recognizer. The qualitative findings also resist a single optimal mode: selectable gestures can reduce setup decisions, custom gestures can improve fit and memorability, and alternatives matter when fatigue or context changes. Adversarial assistance carries these personalized bindings through an unchanged image-recognition pathway; the evaluation estimates the complete pipeline rather than a separately isolated perturbation effect.

\subsection{Integration and Why Image Mediation Matters}

AdaptaFace is intended for systems in which a developer can intercept camera frames before recognition but cannot retrain or replace the downstream recognizer or its command mapping. Integration requires access to that image stream, a differentiable proxy with a compatible target space, and a registration layer mapping user-compatible gestures to internal targets. Proxy/recognizer pairings should be validated rather than assumed to transfer, and facial images and personalized templates should remain local where possible.

Trajectory matching alone selects an internal target; without perturbation, the recognizer still receives the raw frame and therefore follows the same recognizer-facing path as Baseline. Directly emitting the matched label as a command is a valid alternative when an application exposes an appropriate API, but it bypasses the image recognizer and requires a new event source. Image mediation addresses the complementary setting in which the existing image recognizer is the available integration boundary. The choice depends on application access, recognizer coupling, privacy, latency, and the feedback users need to inspect and revise mappings.

These requirements are not specific to facial input. Camera-based hand or body gesture recognizers could use the same pattern when developers can intercept the input stream and construct a proxy with a compatible target space. Beyond vision, one possible direction is fixed class-based recognition: for example, a modality-specific mediation layer could map user-compatible vocalizations to commands supported by a closed-set spoken-command recognizer. Extending the approach would require modality-specific optimization and perceptual constraints, temporal matching, and validation across recognizers. For new user populations, registration and feedback should be designed around their abilities and preferences rather than reuse the facial gesture repertoire evaluated here.

\subsection{Scope, Limitations, and Future Work}

The online study used the same OpenFace-based AU model for perturbation generation and post-assistance scoring, so its gains characterize an aligned proxy/evaluator setting. Offline validation showed that the mechanism could also be instantiated with OpenGraphAU, but OpenGraphAU-to-OpenFace transfer was substantially weaker. Targeted transfer is known to be difficult, although ensemble, feature-space, and intermediate-level objectives offer possible improvement strategies~\cite{liu2017delving,inkawhich2019feature,wang2021admix,li2023ilpd}. These results define a current boundary rather than evidence of universal black-box portability.

The study compared the complete Baseline and AdaptaFace pipelines and did not isolate matching, target selection, and perturbation generation through a component ablation. Its remote deployment also did not systematically manipulate pose, lighting, occlusion, webcam quality, or within-person change. Future work should benchmark matching and mediation across those conditions and recognizer pairings, examine longer-term changes in fatigue and gesture use, and compare image mediation with direct triggering where both architectures are available.

\section{Conclusion}

We presented AdaptaFace, a facial interaction system that treats personalization as input-side accessibility mediation. The complete pipeline improved task performance, perceived control, usability, and workload relative to Baseline, while extension modes broadened participants' usable gesture vocabularies. Together, the results show how feasible, memorable movements can be carried through an unchanged AU-based recognition pathway without retraining. The present evidence remains bounded to proxy-compatible target spaces and does not isolate every pipeline component, but it establishes input-side mediation as a promising direction for more user-compatible facial interaction.

\begin{acks}
We thank all participants for their time and valuable contributions, and the University of Bristol for supporting this research.
\end{acks}

\bibliographystyle{ACM-Reference-Format}
\bibliography{sample-base}

\appendix

\section{AdaptaFace Implementation Details}
\label{app:implementation_details}
\subsection{PGD-style Perturbation Optimization}
\label{app:PGD}

AdaptaFace generates adversarial assistance using a PGD-style targeted optimization procedure with a random start. Given an input frame $x$ and a target recognizer output $y^*$, the system iteratively updates a perturbation $\delta$ to reduce the proxy-model loss for $y^*$ while constraining the perturbation magnitude within a bounded $L_\infty$ norm ball. This yields a visually subtle but effective modification of the input that steers recognition toward the intended target.

At each iteration, AdaptaFace computes the gradient of the targeted loss with respect to the current perturbed input, updates the perturbation in the loss-reducing direction, and projects the result back into the feasible perturbation set. We use only a small number of iterations in practice to preserve real-time performance. Panel E of Figure~\ref{fig:core_mechanism} shows a representative perturbation example; for visibility, the perturbation visualization is magnified.

\begin{algorithm}
  \caption{AdaptaFace perturbation engine (per frame)}
  \label{alg:adaptaface}
  \begin{algorithmic}[1]
    \Require frame $x$, target output $y^{*}$, proxy model $f_{\text{proxy}}$, step size $\alpha$, bound $\epsilon$, iterations $K$
    \State $\delta \gets \mathrm{clip}_{[0,1]}(x+\eta)-x$, where $\eta \sim \mathcal{U}([-\epsilon,\epsilon])$
    \For{$k = 1$ to $K$}
      \State $g \gets \nabla_x \mathcal{L}(f_{\text{proxy}}(x+\delta), y^{*})$
      \State $\delta \gets \delta - \alpha \cdot \mathrm{sign}(g)$
      \State $\delta \gets \mathrm{Proj}_{\|\delta\|_\infty \le \epsilon,\,x+\delta\in[0,1]}(\delta)$
    \EndFor
    \State \Return $x + \delta$
  \end{algorithmic}
\end{algorithm}

\subsection{Target Selection in Matched Modes}

In the matched modes, AdaptaFace assigns a target recognizer output during registration rather than during runtime interaction. Let $\mathcal{G}=\{g_1,\dots,g_N\}$ denote the candidate set. In Matched Enhancement, $\mathcal{G}$ corresponds to the set of interface-supported gestures. In Matched Extension, $\mathcal{G}$ corresponds to the set of available recognizer outputs considered as candidate mediation targets for user-defined gestures.

Given a registered gesture sample or sequence $X^{\mathrm{reg}}$, AdaptaFace performs registration-time candidate evaluation by estimating how well the assisted registration input aligns with each candidate $g_i \in \mathcal{G}$. Let $c_i(X^{\mathrm{reg}})$ denote the downstream recognizer's confidence score for candidate $g_i$. The matched target is selected as
\begin{equation}
g^{*} = \arg\max_{g_i \in \mathcal{G}} c_i(X^{\mathrm{reg}}).
\end{equation}

The selected target is accepted only if its confidence exceeds a minimum threshold $\tau_c$:
\begin{equation}
c_{g^{*}}(X^{\mathrm{reg}}) \ge \tau_c.
\end{equation}
Optionally, AdaptaFace also applies a confidence-margin criterion:
\begin{equation}
c_{g^{*}}(X^{\mathrm{reg}}) - c_{g^{(2)}}(X^{\mathrm{reg}}) \ge \tau_m,
\end{equation}
where $g^{(2)}$ is the second-highest-scoring candidate. If these criteria are not satisfied, no matched target is assigned.

To improve robustness during registration, AdaptaFace can aggregate confidence scores across multiple frames in the recorded sample or require the same candidate to remain top-ranked across a short temporal window before confirming the match.

\subsection{Runtime Gesture Detection}
\label{app:runtime_detection}

Before applying assistance, AdaptaFace first determines whether the user's ongoing facial motion corresponds to one of the registered gestures. Let $\mathcal{R}=\{r_1,\dots,r_M\}$ denote the set of registered gesture trajectories. Each registered gesture $r_j$ is associated with a target recognizer output determined during registration, either predefined in the fixed modes or selected through matching in the matched modes. Each incoming frame is represented by an unnormalized eight-dimensional AU activation vector $\mathbf{a}_t$. At runtime, the system maintains a short causal sliding window
\begin{equation}
A_t = \{\mathbf{a}_{t-W+1}, \dots, \mathbf{a}_t\},
\end{equation}
where $W$ is the window length. Writing the aligned stored trajectory as $r_j=\{\mathbf{r}_{j,1},\dots,\mathbf{r}_{j,W}\}$, the matcher computes

\begin{equation}
D_j(A_t,r_j)=\frac{1}{W}\sum_{i=1}^{W}\left\lVert\mathbf{a}_{t-W+i}-\mathbf{r}_{j,i}\right\rVert_2,
\end{equation}
followed by the distance-derived similarity
\begin{equation}
s_j(A_t,r_j)=\frac{1}{1+D_j(A_t,r_j)}.
\end{equation}
The best-matching registered gesture is defined as
\begin{equation}
r^{*} = \arg\max_{r_j \in \mathcal{R}} s_j(A_t, r_j),
\end{equation}
with equal scores resolved deterministically by template identifier.

A gesture is considered detected only if the best match exceeds a minimum similarity threshold:
\begin{equation}
s_{r^{*}}(A_t, r^{*}) \ge \tau_{\text{match}}.
\end{equation}
The matcher exposes $T$ as a configurable stability requirement and can additionally require the same gesture to remain top-ranked and above threshold for $T$ consecutive windows before confirmation. The study configuration used $W=1$, $\tau_{\mathrm{match}}=2/3$, and $T=3$, with no AU-vector normalization. The similarity threshold is equivalent to a Euclidean-distance threshold of $0.5$. Thus, confirmation required three consecutive above-threshold frame-level matches to the same template, suppressing transient matches.

Rather than waiting for a full gesture sequence to complete, AdaptaFace uses short-window matching to detect gesture onset online. Once a registered gesture is detected, the compatibility layer applies adversarial assistance frame by frame to subsequent input during interaction, steering recognition toward the target previously assigned to that registered gesture.

\section{Extended User-Study Materials}
\label{app:extended_userstudy}
\subsection{Statistical Model Specifications}
\label{app:stats}

\subsubsection{Modeling Setup}

We analyzed repeated-measures quantitative data using two complementary model families. CR was summarized as the within-block proportion of required action repetitions completed. For inferential analysis, full-block completion, a binary outcome indicating whether all required repetitions in a task block were completed, was analyzed using a binomial generalized estimating equation (GEE) with a logit link and participant-level clustering. Continuous repeated-measures outcomes were analyzed using linear mixed-effects models (LMMs) with participant-level random intercepts. These outcomes included task completion time (TCT), success-normalized task time (SNT), ease of use, and perceived control.

Technology (Baseline vs.\ AdaptaFace), Task, and Personalization Mode were entered as fixed effects in the repeated-measures models. Because no Technology $\times$ Task or Technology $\times$ Mode interaction reached significance in the final analyses, the main paper focuses on Technology main effects, while task- and mode-level patterns are interpreted descriptively. System-level SUS and NASA Raw TLX were collected once per technology condition and are therefore summarized descriptively in the main paper rather than modeled as repeated-measures outcomes.

\subsubsection{Objective Performance Models}

\paragraph{Full-Block Completion.}
The binary full-block completion indicator derived from CR was modeled using a binomial GEE with a logit link:
\[
\begin{aligned}
\Pr(\mathrm{FBC}_{ij}=1)
&=
\operatorname{logit}^{-1}
\big(
\beta_0 + \beta_T \mathrm{Technology}_{ij} \\
&\quad
+ \beta_{Ta} \mathrm{Task}_{ij}
+ \beta_M \mathrm{Mode}_{ij}
\big),
\end{aligned}
\]
where $i$ indexes participants and $j$ indexes task blocks.

\paragraph{Task Completion Time (TCT).}
Task completion time was modeled using a linear mixed-effects model:
\[
\mathrm{TCT}_{ij}
=
\beta_0
+ \beta_T \mathrm{Technology}_{ij}
+ \beta_{Ta} \mathrm{Task}_{ij}
+ \beta_M \mathrm{Mode}_{ij}
+ u_i + \epsilon_{ij},
\]
where $u_i$ is a participant-level random intercept.

\paragraph{Success-Normalized Task Time (SNT).}
Success-normalized task time was modeled analogously:
\[
\mathrm{SNT}_{ij}
=
\beta_0
+ \beta_T \mathrm{Technology}_{ij}
+ \beta_{Ta} \mathrm{Task}_{ij}
+ \beta_M \mathrm{Mode}_{ij}
+ u_i + \epsilon_{ij}.
\]

\subsubsection{Subjective Rating Models}

\paragraph{Ease and Perceived Control.}
Ease and perceived control ratings were analyzed using linear mixed-effects models with the same fixed- and random-effects structure:
\[
Y_{ij}
=
\beta_0
+ \beta_T \mathrm{Technology}_{ij}
+ \beta_{Ta} \mathrm{Task}_{ij}
+ \beta_M \mathrm{Mode}_{ij}
+ u_i + \epsilon_{ij},
\]
where $Y_{ij}$ denotes either Ease or Control.

\paragraph{System-Level SUS and NASA Raw TLX.}
SUS and NASA Raw TLX were collected once per participant for each technology condition rather than once per task block. For this reason, they are treated as system-level comparisons in the main paper rather than repeated-measures model outcomes.

\subsubsection{Fixed-Effect Tables}

Table~\ref{tab:appendix_cr_simple} reports the GEE fixed effects for full-block completion.
Tables~\ref{tab:appendix_tct_simple} and \ref{tab:appendix_snt_simple} report the LMM fixed effects for TCT and SNT.
Tables~\ref{tab:appendix_ease_simple} and \ref{tab:appendix_control_simple} report the LMM fixed effects for Ease and Control.
Across these tables, the reference levels are Baseline, Click, and Fixed Extension.

\begin{table*}[t]
\centering
\caption{GEE fixed effects for full-block completion (logit link). Only interpretable fixed-effect terms are reported.}
\label{tab:appendix_cr_simple}
\small
\begin{tabular}{lrrrrrr}
\toprule
Term & Estimate & StdErr & $z$ & $p$ & CI$_{low}$ & CI$_{high}$ \\
\midrule
Task: Swipe & -0.326 & 0.132 & -2.461 & 0.014 & -0.585 & -0.066 \\
Task: Zoom In & -0.346 & 0.131 & -2.637 & 0.008 & -0.604 & -0.089 \\
Task: Zoom Out & -0.066 & 0.135 & -0.493 & 0.622 & -0.330 & 0.198 \\
Mode: Matched Extension & 0.588 & 0.133 & 4.410 & $<.001$ & 0.327 & 0.849 \\
Mode: Fixed Enhancement & 1.230 & 0.133 & 9.242 & $<.001$ & 0.970 & 1.491 \\
Mode: Matched Enhancement & 1.153 & 0.133 & 8.697 & $<.001$ & 0.893 & 1.413 \\
Technology: AdaptaFace & 1.116 & 0.094 & 11.864 & $<.001$ & 0.931 & 1.300 \\
\bottomrule
\end{tabular}
\end{table*}

\begin{table*}[t]
\centering
\caption{LMM fixed effects for Task Completion Time (seconds). Only interpretable fixed-effect terms are reported.}
\label{tab:appendix_tct_simple}
\small
\begin{tabular}{lrrrrrr}
\toprule
Term & Estimate & StdErr & $z$ & $p$ & CI$_{low}$ & CI$_{high}$ \\
\midrule
Technology: AdaptaFace & -3.183 & 0.402 & -7.924 & $<.001$ & -3.971 & -2.395 \\
Task: Swipe & 0.854 & 0.655 & 1.304 & 0.192 & -0.431 & 2.138 \\
Task: Zoom In & 0.827 & 0.636 & 1.300 & 0.194 & -0.420 & 2.074 \\
Task: Zoom Out & 0.399 & 0.665 & 0.600 & 0.549 & -0.905 & 1.704 \\
Mode: Matched Extension & -3.879 & 0.641 & -6.052 & $<.001$ & -5.136 & -2.622 \\
Mode: Fixed Enhancement & -0.990 & 0.657 & -1.507 & 0.132 & -2.278 & 0.299 \\
Mode: Matched Enhancement & -1.453 & 0.659 & -2.206 & 0.027 & -2.744 & -0.163 \\
\bottomrule
\end{tabular}
\end{table*}

\begin{table*}[t]
\centering
\caption{LMM fixed effects for Success-Normalized Task Time (seconds per successful action). Only interpretable fixed-effect terms are reported.}
\label{tab:appendix_snt_simple}
\small
\begin{tabular}{lrrrrrr}
\toprule
Term & Estimate & StdErr & $z$ & $p$ & CI$_{low}$ & CI$_{high}$ \\
\midrule
Technology: AdaptaFace & -12.157 & 0.963 & -12.626 & $<.001$ & -14.046 & -10.269 \\
Task: Swipe & -0.420 & 1.464 & -0.287 & 0.774 & -3.289 & 2.448 \\
Task: Zoom In & -6.441 & 1.445 & -4.459 & $<.001$ & -9.272 & -3.610 \\
Task: Zoom Out & 5.563 & 1.526 & 3.646 & $<.001$ & 2.571 & 8.556 \\
Mode: Matched Extension & -10.378 & 1.456 & -7.126 & $<.001$ & -13.232 & -7.525 \\
Mode: Fixed Enhancement & -7.334 & 1.488 & -4.933 & $<.001$ & -10.250 & -4.418 \\
Mode: Matched Enhancement & -5.900 & 1.490 & -3.959 & $<.001$ & -8.830 & -2.970 \\
\bottomrule
\end{tabular}
\end{table*}

\begin{table*}[t]
\centering
\caption{LMM fixed effects for Ease ratings (1--5). Only interpretable fixed-effect terms are reported.}
\label{tab:appendix_ease_simple}
\small
\begin{tabular}{lrrrrrr}
\toprule
Term & Estimate & StdErr & $z$ & $p$ & CI$_{low}$ & CI$_{high}$ \\
\midrule
Technology: AdaptaFace & 0.674 & 0.051 & 13.118 & $<.001$ & 0.573 & 0.775 \\
Task: Swipe & 0.056 & 0.050 & 1.115 & 0.265 & -0.042 & 0.154 \\
Task: Zoom In & -0.061 & 0.049 & -1.241 & 0.215 & -0.158 & 0.036 \\
Task: Zoom Out & 0.145 & 0.051 & 2.859 & 0.004 & 0.045 & 0.245 \\
Mode: Matched Extension & 0.529 & 0.051 & 10.377 & $<.001$ & 0.428 & 0.629 \\
Mode: Fixed Enhancement & -0.093 & 0.052 & -1.788 & 0.074 & -0.195 & 0.009 \\
Mode: Matched Enhancement & -0.062 & 0.052 & -1.192 & 0.233 & -0.164 & 0.039 \\
\bottomrule
\end{tabular}
\end{table*}

\begin{table*}[t]
\centering
\caption{LMM fixed effects for Perceived Control ratings (1--5). Only interpretable fixed-effect terms are reported.}
\label{tab:appendix_control_simple}
\small
\begin{tabular}{lrrrrrr}
\toprule
Term & Estimate & StdErr & $z$ & $p$ & CI$_{low}$ & CI$_{high}$ \\
\midrule
Technology: AdaptaFace & 1.124 & 0.055 & 20.478 & $<.001$ & 1.016 & 1.231 \\
Task: Swipe & 0.045 & 0.054 & 0.824 & 0.410 & -0.061 & 0.152 \\
Task: Zoom In & -0.006 & 0.054 & -0.108 & 0.914 & -0.111 & 0.100 \\
Task: Zoom Out & 0.423 & 0.056 & 7.528 & $<.001$ & 0.313 & 0.533 \\
Mode: Matched Extension & 0.503 & 0.055 & 9.119 & $<.001$ & 0.395 & 0.612 \\
Mode: Fixed Enhancement & -0.331 & 0.056 & -5.913 & $<.001$ & -0.440 & -0.222 \\
Mode: Matched Enhancement & -0.197 & 0.056 & -3.548 & $<.001$ & -0.307 & -0.087 \\
\bottomrule
\end{tabular}
\end{table*}

\subsection{Formative-Study Interview Procedure}
\label{sec:formativestudy}
The main paper reports the interview sequence and analysis purpose in Section~\ref{sec:formative}. Sessions were conducted remotely by video conference, recorded with consent, and structured to preserve participants' ability to pause or withdraw.
\subsection{Extended Design Principles from the Formative Study}
\label{appendix:extended_formative}

Based on the formative findings, we derived six design principles for inclusive facial-gesture personalization. In the main paper, we summarize these principles briefly; here we provide a slightly fuller description of how each informed the design of AdaptaFace.

\paragraph{DP1. Support user-compatible gestures rather than only canonical templates.}
Personalization should accommodate gestures that users can comfortably and reliably perform, even when these differ from recognizer-defined expressions. This follows from participants' accounts that subtle, asymmetric, or personally habitual movements were often more feasible than standardized facial gestures.

\paragraph{DP2. Minimize user burden during personalization.}
Personalization should reduce repeated demonstration, calibration, and setup effort, especially for users whose movement quality varies across posture, fatigue, or day-to-day condition.

\paragraph{DP3. Enable user-defined gesture vocabularies.}
Users should be able to bind their own feasible and meaningful facial actions to system commands, rather than selecting only from a closed predefined set. This supports interaction that reflects users' existing expressive habits instead of forcing all input into a fixed recognizer vocabulary.

\paragraph{DP4. Preserve user control over personalization.}
Assistance should remain legible and user-governed, allowing users to understand and shape how their gestures are interpreted. Personalization should therefore support explicit gesture--command mappings and avoid opaque adaptation that removes user agency.

\paragraph{DP5. Preserve compatibility with existing recognition pipelines.}
To lower deployment barriers, personalization should work with off-the-shelf recognizers rather than requiring wholesale replacement or retraining of downstream models. This also supports broader applicability across heterogeneous recognition pipelines, rather than binding personalization to a single recognizer architecture.

\paragraph{DP6. Support low-effort and socially acceptable interaction.}
The system should accommodate gestures that are low-fatigue, low-amplitude, and appropriate for everyday settings, including public or shared environments. Inclusive personalization should therefore optimize not only for recognizability, but also for comfort, sustainability, and social acceptability.

These principles point toward a different locus of personalization: a mediating mechanism at the input level that preserves users' own expressive habits while helping existing recognizers interpret them more reliably.

\subsection{Remote Study Setup}
\begin{figure*}[t]
  \centering
  \includegraphics[width=\textwidth]{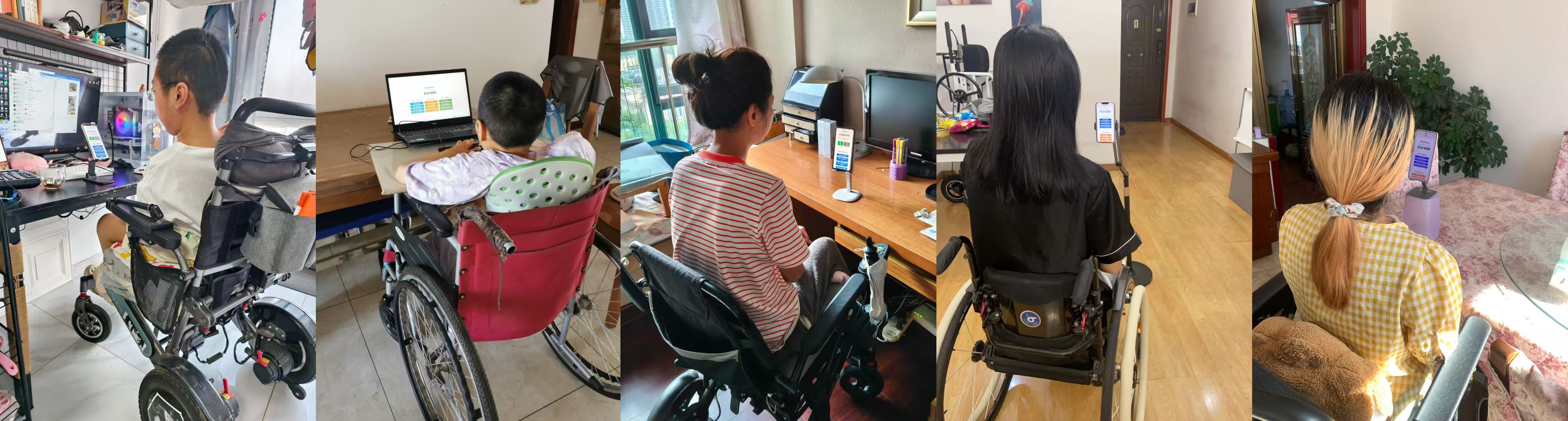}
  \caption{Representative remote-study setups. Participants completed the evaluation using their own devices and webcams in varied postures and home environments.}
  \Description{Five rear-view photographs show participants using AdaptaFace remotely on their own computers and mobile devices from varied seated positions and home environments.}
  \label{fig:setup}
\end{figure*}

To illustrate the evaluation context, Figure~\ref{fig:setup} shows representative remote-study setups.

\subsection{Task-Level Performance}
Table~\ref{tab:perf_by_task} provides the task-level performance breakdown for the four interaction tasks.
\begin{table*}[t]
\centering
\caption{Task-level performance breakdown (mean (SD)). Zoom tasks are reported separately.}
\small
\setlength{\tabcolsep}{4.5pt}
\begin{tabular}{%
  p{1.6cm}
  >{\centering\arraybackslash}p{1.35cm} >{\centering\arraybackslash}p{1.35cm}
  >{\centering\arraybackslash}p{1.35cm} >{\centering\arraybackslash}p{1.35cm}
  >{\centering\arraybackslash}p{1.35cm} >{\centering\arraybackslash}p{1.35cm}
  >{\centering\arraybackslash}p{1.35cm} >{\centering\arraybackslash}p{1.35cm}}
\toprule
Task & \multicolumn{2}{c}{CR\,\%} & \multicolumn{2}{c}{TCT\,s} & \multicolumn{2}{c}{SNT\,s} & \multicolumn{2}{c}{Conf.} \\
\cmidrule(lr){2-3} \cmidrule(lr){4-5} \cmidrule(lr){6-7} \cmidrule(lr){8-9}
 & Baseline & AdaptaFace & Baseline & AdaptaFace & Baseline & AdaptaFace & Baseline & AdaptaFace \\
\midrule
Click    & 70.0 (41.1) & 95.5 (18.2) & 11.5 (7.4)  & 8.4 (4.7)  & 17.6 (19.2) & 8.8 (5.0)  & 0.78 (0.37) & 0.92 (0.24) \\
Swipe    & 55.5 (46.9) & 87.0 (29.0) & 17.6 (12.0) & 13.3 (8.7) & 21.8 (23.8) & 18.2 (24.4) & 0.56 (0.45) & 0.87 (0.30) \\
Zoom In  & 52.1 (43.9) & 89.1 (28.1) & 24.0 (27.1) & 18.3 (10.2) & 96.1 (292.9) & 20.7 (11.9) & 0.62 (0.42) & 0.88 (0.28) \\
Zoom Out & 61.1 (45.6) & 90.3 (25.6) & 13.1 (9.2)  & 13.4 (8.4)  & 14.2 (14.2)  & 16.3 (14.5) & 0.64 (0.45) & 0.87 (0.28) \\
\bottomrule
\end{tabular}

\label{tab:perf_by_task}
\end{table*}

\subsection{Ease and Control Ratings}
Table~\ref{tab:ease-control-all} reports subjective ease and control ratings overall, by task, and by personalization mode.
\begin{table*}[t]
\centering
\small
\caption{Ease and control ratings (1--5; higher is better).}
\label{tab:ease-control-all}

\begin{subtable}[t]{0.28\linewidth}
\centering
\caption{Overall}
\label{tab:ease-control-overall}
\begin{tabular}{lrr}
\toprule
 & Ease & Control \\
\midrule
Baseline   & 3.70 & 3.15 \\
AdaptaFace & 4.35 & 4.27 \\
\bottomrule
\end{tabular}
\end{subtable}\hfill
\begin{subtable}[t]{0.32\linewidth}
\centering
\caption{By task}
\label{tab:ease-control-task}
\begin{tabular}{lrrrr}
\toprule
Task & \multicolumn{2}{c}{Ease} & \multicolumn{2}{c}{Control} \\
\cmidrule(lr){2-3}\cmidrule(lr){4-5}
 & Base & Ada & Base & Ada \\
\midrule
Click     & 3.55 & 4.13 & 2.82 & 3.99 \\
Swipe     & 3.66 & 4.39 & 3.11 & 4.26 \\
Zoom In   & 3.62 & 4.34 & 3.05 & 4.27 \\
Zoom Out  & 3.78 & 4.38 & 3.63 & 4.53 \\
\bottomrule
\end{tabular}
\end{subtable}\hfill
\begin{subtable}[t]{0.36\linewidth}
\centering
\caption{By mode}
\label{tab:ease-control-mode}
\begin{tabular}{lrrrr}
\toprule
Mode & \multicolumn{2}{c}{Ease} & \multicolumn{2}{c}{Control} \\
\cmidrule(lr){2-3}\cmidrule(lr){4-5}
 & Base & Ada & Base & Ada \\
\midrule
Fixed Enhancement   & 3.58 & 4.22 & 2.96 & 4.04 \\
Matched Enhancement & 3.62 & 4.28 & 2.98 & 4.14 \\
Fixed Extension     & 3.75 & 4.41 & 3.22 & 4.33 \\
Matched Extension   & 3.86 & 4.51 & 3.44 & 4.55 \\
\bottomrule
\end{tabular}
\end{subtable}
\end{table*}

\subsection{Gesture Vocabulary Inventory}
Table~\ref{tab:user-defined-gestures-completion-tech-delta} summarizes the user-defined gestures introduced in the extension-oriented modes.
\begin{table*}[t]
\centering
\caption{Gesture-level breakdown of user-defined gestures introduced in the extension-oriented modes, with descriptive completion rates by technology and improvement ($\Delta$).}
\setlength{\tabcolsep}{4pt}
\begin{tabular}{%
  p{3.2cm}  
  p{3.3cm}  
  p{2.9cm}  
  >{\centering\arraybackslash}p{0.8cm} 
  >{\centering\arraybackslash}p{1.5cm} 
  >{\centering\arraybackslash}p{1.5cm} 
  >{\centering\arraybackslash}p{1.1cm} 
}
\toprule
Action (English) & Interactions & Participants & Count & Baseline (\%) & AdaptaFace (\%) & $\Delta$ (\%) \\
\midrule
Right Brow Raise     & Swipe & EP2 & 1 & 60.0 & 100.0 & 40.0 \\
Eye Widen            & Zoom In & EP16 & 1 & 60.0 & 100.0 & 40.0 \\
Half-squinted Eyes   & Zoom Out & EP13 & 1 & 0.0 & 100.0 & 100.0 \\
Close Eyes           & Swipe, Click, Zoom In & EP12, EP16, EP3, EP5, EP8 & 11 & 14.5 & 66.0 & 51.5 \\
Close Left Eye       & Swipe, Zoom In/Out & EP10, EP14, EP5, EP9 & 7 & 22.9 & 90.0 & 67.1 \\
Close Right Eye      & Swipe, Click, Zoom In/Out & EP10, EP11, EP14, EP2, EP3, EP6, EP8 & 11 & 40.0 & 83.6 & 43.6 \\
Eye Look Left        & Swipe & EP9 & 1 & 0.0 & 100.0 & 100.0 \\
Eye Look Up          & Click, Zoom Out & EP12, EP13, EP6 & 4 & 25.0 & 75.0 & 50.0 \\
Eye Look Down        & Zoom Out & EP13 & 1 & 40.0 & 100.0 & 60.0 \\
Puffing up cheeks    & Zoom In/Out & EP13, EP2, EP3, EP6 & 5 & 20.0 & 100.0 & 80.0 \\
Mouth Widen          & Click, Zoom In & EP1, EP10, EP13, EP8, EP15 & 6 & 38.3 & 100.0 & 61.7 \\
Lip Press            & Swipe, Click, Zoom In/Out & EP13, EP9, EP2, EP3, EP6 & 6 & 33.3 & 96.7 & 63.4 \\
Pout                 & Swipe, Click, Zoom In & EP1, EP11, EP13, EP15, EP16, EP3, EP8, EP9 & 13 & 33.8 & 91.7 & 57.9 \\
Stretch Left Lip      & Zoom In & EP7 & 1 & 0.0 & 60.0 & 60.0 \\
Stretch Right Lip     & Zoom In & EP7 & 1 & 20.0 & 100.0 & 80.0 \\
Raise Left Lip Corner& Swipe, Click & EP11, EP3, EP9 & 5 & 35.0 & 88.0 & 53.0 \\
Raise Right Lip Corner & Click & EP14, EP9, EP2 & 4 & 15.0 & 70.0 & 55.0 \\
Smile (teeth)        & Swipe & EP1, EP13, EP4 & 3 & 26.7 & 86.7 & 60.0 \\
Depress Mouth Corners& Click & EP3 & 1 & 100.0 & 100.0 & 0.0 \\
Tongue Out           & Zoom In, Zoom Out & EP11, EP9, EP4 & 4 & 25.0 & 90.0 & 65.0 \\
\bottomrule
\end{tabular}
\label{tab:user-defined-gestures-completion-tech-delta}
\end{table*}

\section{Additional Offline Validation Examples}
\label{app:offline_validation_examples}
\begin{figure*}[t]
  \centering
  \includegraphics[width=\textwidth]{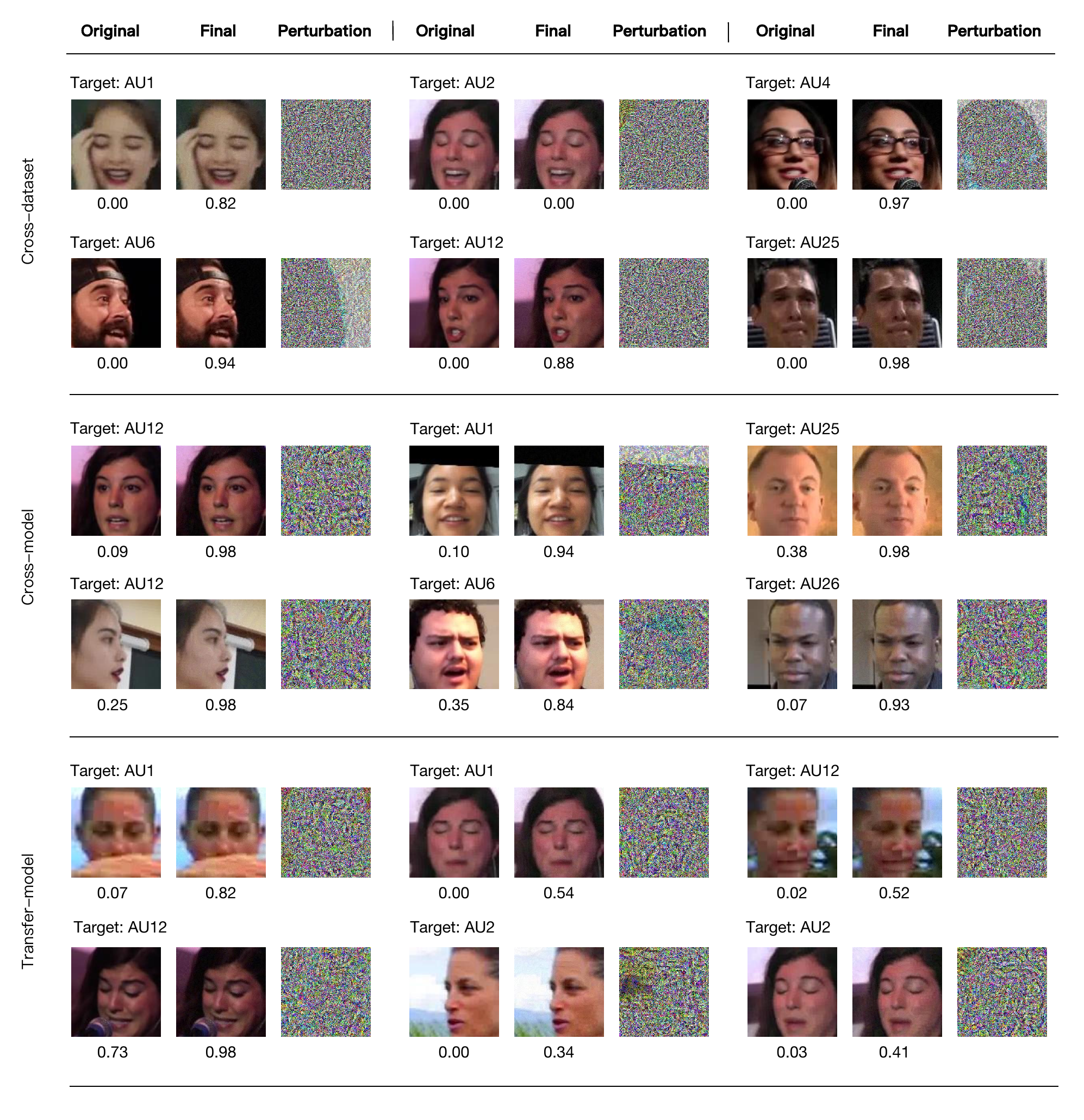}
  \caption{Additional qualitative examples from the three offline validation settings: cross-dataset, cross-model, and model transfer. Each triplet shows the original input, the assisted output, and a normalized perturbation visualization.}
  \Description{Additional image triplets compare original inputs, assisted outputs, and magnified perturbations for cross-dataset, cross-model, and model-transfer validation cases.}
  \label{fig:validation_imgs}
\end{figure*}

\end{document}